\documentclass{jkas}

\def\year{2014} 
\def\month{May} 
\def\volume{00} 
\def\issue{0} 
\def\beginpage{1} 
\def\endpage{99} 
\def\received{February 30, 2014} 
\def\accepted{February 31, 2014} 
\date{Received \received; accepted \accepted}

\title{
Large SDSS Quasar Groups and Their Statistical Significance
}

\author[1]{Changbom~Park}
\author[1]{Hyunmi~Song}
\author[2]{Maret~Einasto}
\author[3,4]{Heidi~Lietzen}
\author[5]{Pekka~Hein{\"a}m{\"a}ki}

\affil[1]{School of Physics, Korea Institute for Advanced Study, Heogiro 85, Seoul 130-722, Korea; \email{cbp@kias.re.kr, hmsong@kias.re.kr}}
\affil[2]{Tartu Observatory, 61602 Toravere, Estonia; \email{maret@to.ee}}
\affil[3]{Instituto de Astrof{\'i}sica de Canarias, E-38205 La Laguna, Tenerife, Spain; \email{heidi@aai.ee}}
\affil[4]{Universidad de La Laguna, Dept. Astrof{\'i}sica, E-38206 La Laguna, Tenerife, Spain}
\affil[5]{Tuorla Observatory, University of Turku, V{\"a}is{\"al}{\"a}ntie 20, Piikki{\"o}, Finland; \email{pekheina@utu.fi}}

\def\corrauthor{
H. Song
}

\def\runningauthor{
Park et al.
}

\def\runningtitle{
Large Quasar Groups
}

\def\keywords{
large-scale structure of universe -- cosmology: observations -- quasars: general
}

\def\abstracttext{
We use a volume-limited sample of quasars in the Sloan Digital Sky Survey (SDSS)
DR7 quasar catalog
to identify quasar groups and address their statistical significance.
This quasar sample has a uniform selection function on the sky and nearly 
a maximum possible contiguous volume that can be drawn from the DR7 catalog. 
Quasar groups are identified by using the Friend-of-Friend algorithm with
a set of fixed comoving linking lengths.
We find that the richness distribution of the richest 100 quasar groups or the size distribution
of the largest 100 groups
are statistically equivalent with those of randomly-distributed points
with the same number density and sky coverage
when groups are identified with the linking length of 70 $h^{-1}$Mpc. 
It is shown that the large-scale structures like
the huge Large Quasar Group (U1.27) reported by \citet{clowes-etal2013} 
can be found with high probability even if quasars have no physical clustering,
and does not challenge the initially homogeneous cosmological models.
Our results are statistically more reliable than those of \citet{nadathur2013}, where the test was made only for the largest quasar group.
It is shown that the linking length should be smaller than 50 $h^{-1}$Mpc 
in order for the quasar groups identified in the DR7 catalog
not to be dominated by associations of quasars grouped by chance.
We present 20 richest quasar groups identified with the linking length of 
70 $h^{-1}$Mpc for further analyses.
}

\begin{document}
\jkashead 


\section{Introduction\label{sec:intro}}
The size and mass of large-scale structures
\footnote{
We use the word `large-scale structure' as a general name for structures 
larger than galaxy clusters. 
In this terminology the large-scale structures of the universe are classified 
into high-density and low-density large-scale structures. The high-density 
large-scale structures are classified into categories of superclusters, 
filaments, and walls depending on their internal density and morphology. 
The low-density large-scale structures are voids, and classified 
into bubbles, and tunnels depending on their morphology. In this scheme
all the names that have been used to describe `large' structures in the universe
are unified.
}
(hereafter LSS) in the universe 
have been often used as measures of amplitude of 
density fluctuations on large scales. The CfA Great Wall discovered by 
\citet{delapparent-etal1986} in the CfA2 survey and the Sloan Great Wall 
in the SDSS survey \citep{gott-etal2005} cast doubt on the SCDM model
\citep{geller_huchra1989} and the $\Lambda$CDM model \citep{sheth_diaferio2011},
respectively.
It has been argued that the universe did not have enough time for such 
large structures to form by the present epoch in these cosmological 
models, and that the gravitational instability theory for structure formation
or the Cosmological Principle adopting large-scale homogeneity 
may not be valid \citep[see also][]{clowes-etal2012,horvath-etal2013}. 

On the other hand, a strong support for the Cosmological Principle comes from the smooth 
cosmic microwave background radiation, which shows anisotropies that agrees astonishingly well
with those of the initially homogeneous isotropic $\Lambda$CDM model \citep{clifton-etal2012,planckcollabo2014}.
The fluctuation of the galaxy number density on large scales is also found 
to agree with the prediction of the $\Lambda$CDM model \citep{scrimgeour-etal2012}.

It is expected that LSS of the universe contain a wealth of information on
primordial density fluctuations and the history of their growth. This is because
they represent density fluctuations on linear scales. 
It is then very important to rigorously study the physical properties and cosmological 
meaning of observed superstructures of the universe \citep{park1990,park-etal2012, 
einasto-etal2011,einasto-etal2014}.

Recently, \citet{clowes-etal2013} reported a finding of a very large quasar group
in the SDSS DR7 quasar catalog \citep{schneider-etal2010}
with characteristic size of $\sim$350 $h^{-1}$Mpc, longest dimension 
of $\sim$870 $h^{-1}$Mpc, membership of 73 quasars, and mean redshift
${\bar z}=1.27$.
They claimed that the quasar group, named U1.27, is the largest `structure'
currently known in the universe.
They argued that their finding might raise a question on validity 
of the large-scale homogeneity assumption, which is, in the form of Cosmological
Principle, one of the fundamental assumptions of the standard cosmology.
Their claim falls on the same kind of view applied to CfA and Sloan Great
Walls.

However, \citet{nadathur2013} pointed out that the statistical significance of the huge quasar group has not been 
properly studied in \citet{clowes-etal2013}, and that Clowes et al.'s 
claim on the cosmological implication of existence of the large quasar
group needs to be examined quantitatively. Nadathur showed that the algorithm used to 
identify the quasar groups frequently finds such large-size structures
even in homogeneous simulations of a Poisson point process with the same
number density as the quasar catalog. They concluded that Clowes et al.'s
interpretation of U1.27 as a physical `structure' is misleading.

Even though it is cosmologically interesting to search for associations
of quasars
\footnote{
In this paper we call the large loose groupings of quasars `quasar groups.'
This terminology can give a misleading idea that they are compact and physically interacting internally
as in galaxy groups. \citet{einasto-etal2014} used `quasar systems' to avoid the confusion.
The name `quasar association' may also be an appropriate name for these loose quasar groups
since the name `stellar association' has been already used in astronomy
for loose groups of stars that are not gravitationally bound.
}
\citep{webster1982,clowes_campusano1991,komberg-etal1996, 
williger-etal2002,clowes-etal2012,clowes-etal2013,einasto-etal2014},
one needs to be careful in interpreting the quasar groups
identified with specific criteria.
This is because the size and richness of LSS depend
sensitively on how they are identified. Evolution of the objects used to 
trace LSS and evolution of LSS themselves 
can also complicate the interpretation of identified LSS.
In particular, the quasar phenomenon has short lifetime
and astrophysics of quasar activity can affect the results
\citep{komberg-etal1996,wold-etal2000,miller-etal2004,
sochting-etal2004,coldwell_lambas2006,hutchings-etal2009, 
lietzen-etal2009,lietzen-etal2011,bradshaw-etal2011,portinari-etal2012, 
trainor_steidel2012,krumpe-etal2013,fanidakis-etal2013, 
shen-etal2013,dipompeo-etal2014,karhunen-etal2014,song-etal2015}.
Quasar samples usually have very low spatial number density, which
can introduce shot noise effects in group identification.
This situation demands the use of
well-defined samples and rigorous statistical tests.

In this paper we identify quasar groups using a volume-limited sample
drawn from the SDSS DR7 quasar catalog that has the maximum contiguous 
angular size. We present the statistical significance of not only the
largest quasar group, but also the quasar groups within a large range of 
richness.  We also address a question whether or not the huge quasar group 
found by Clowes et al. is a statistically-significant object
in terms of multiplicity and size.
We find the critical linking length (LL hereafter) for quasar group 
identification which can result in a catalog dominated by physically clustered groups.

\section{SDSS Quasar Catalog}
We use the fifth edition of the SDSS quasar catalog from SDSS DR7, which is a
compilation of quasars probed in the SDSS-I and SDSS-II quasar survey
\citep{schneider-etal2010}. 
The catalog contains 105783 spectroscopically confirmed quasars
in the sky of area covering approximately $9380\textrm{deg}^2$.
They are brighter than $M_i=-22.0$ (corrected for the Galactic extinction,
and $K$-corrected to $z=2$ in a cosmology with
$H_0=70\textrm{km/s/Mpc}$, $\Omega_m=0.3$, and $\Omega_\Lambda=0.7$)
and have at least one broad emission line with full width at half maximum 
larger than $1000\textrm{km/s}$ or interesting/complex absorption features.
The quasars show a wide redshift distribution from $0.065$ to $5.46$,
but the majority of quasars (76\%) is concentrated below redshift $2$.
They have the apparent 
$i$ magnitude of $14.86<i<22.36$, where the lower limit comes from
the maximum brightness limit of the target selection on quasar candidates
to avoid saturation and cross-talk in the spectra.
The catalog does not include several classes of active galactic nuclei
such as Type 2 quasars, Seyfert galaxies, and BL Lacertae objects.

Even though the catalog contains a large number of quasars, not all the quasars are suitable
for statistical analyses for several reasons. It is mainly because the quasar target selection
for the follow-up spectroscopy was not done uniformly in its early version
\citep[see][for more details]{richards-etal2006,shen-etal2007}.
\citet{shen-etal2011} provided a parameter, in the $10${\it th} column in their catalog, 
named as `uniform flag' to take the uniformity problem into account.
The catalog of Schneider et al. also has its own uniform flag in the 
35th column. The uniform flag is assigned based on the achievement 
of the target selection algorithm. But Shen et al. made more careful 
evaluation of the selection effects by finding out 
quasars that are uniformly selected as high-$z$ candidates but later 
identified as low-$z$ quasars. Shen et al. has one more class than 
Schneider et al. as follows: $0=$ not in the uniform sample; $1=$ 
uniformly selected and with the galactic extinction-corrected $i$ magnitude
less than 19.1 at $z<2.9$ or 20.2 at $z>2.9$; $2=$ uniformly selected 
by the QSO\_HiZ branch in the algorithm and with measured spectroscopic 
redshift $z<2.9$ and $i>19.1$. 
%
%
By selecting those whose \citet{shen-etal2011}'s uniform flag is $1$,
we construct a uniform sample of 59679 quasars with Galactic extinction corrected
$i$ magnitude less than 19.1.
It rejects about a half of the quasars in the original catalog, mostly at $z<2.9$.

In Figure 1 we plot the quasars in the \citet{schneider-etal2010}'s catalog 
in black. Only the northern hemisphere of the SDSS survey regions is adopted and 
the SDSS survey coordinates $\eta$ and $\lambda$ are used for the plot
as the survey boundaries are more simply defined in this coordinate system.  
The red points in the figure are quasars marked as those of the uniform subsample by Shen et al.

Since the sky distribution of the uniform quasar sample is not perfectly contiguous,
we define a new boundary for our sample by excluding isolated patches and 
trimming the jagged boundaries to make one clean contiguous region.
The green line in Figure 1 defines the region under our study.
The area of the region is 1.754 steradian.
This is nearly the maximum possible contiguous sample can be drawn 
from the DR7 catalog.

Figure 2 shows the uniform-sample quasars located within our mask in the 
$i$-band absolute magnitude and redshift plane.
In Figure 3 it can be seen
that the comoving space number density of quasars is 
roughly uniform between 
$z=0.8$ and 1.8, but is significantly higher between $z=0.1$ and 0.8. 

\citet{song-etal2015} has compared the distribution of the SDSS quasars
at $0.4\le z \le 0.7$
with the smooth density field of the galaxies in the SDSS DR12 BOSS survey
and found that the probability of finding a quasar increases monotonically
as the local galaxy number density increases 
with no significant redshift dependence.
This implies that LSS found in the distribution of quasars
should be similar to those in the galaxy distribution.
Then a quasar sample having a constant comoving number density should
be as useful as a volume-limited galaxy sample in finding LSS.

In order to make a volume-limited quasar sample with uniform density 
we apply an absolute magnitude cut line given by 
\begin{equation}
M_{i,{\rm lim}}=0.7216 z^2 -3.505 z -22.278
\end{equation}
in the range of $0.1<z<0.8$. The cut is shown as a blue line in Figure 2.
The resulting subsample has a nearly uniform number density of $1.31\times 10^{-6}
(h^{-1}$Mpc)$^{-3}$ in a wide range of $0.1<z<1.8$ as shown by the red histogram
in Figure 3. The corresponding mean quasar separation is 91.5 $h^{-1}$Mpc.
By varying the absolute magnitude cut one can effectively correct a quasar sample
for the luminosity evolution.
The use of uniform comoving density sample makes us to sample the same kind of
cosmological objects over a wide redshift range.

\begin{figure}
\centering
\includegraphics[width=80mm]{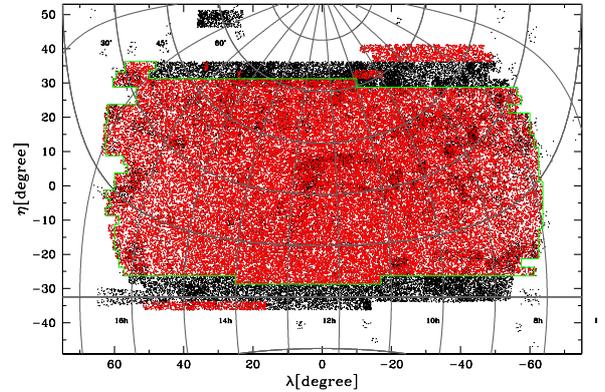}
\caption{Distribution of all the quasars in the original \citet{schneider-etal2010}'s 
catalog (black points) and the quasars in the \citet{shen-etal2011}'s uniform sample 
(red points) in the SDSS survey coordinates. 
A contiguous region defined by the green line is the region under analysis 
in our study.}
\end{figure}

\begin{figure}
\centering
\includegraphics[width=80mm]{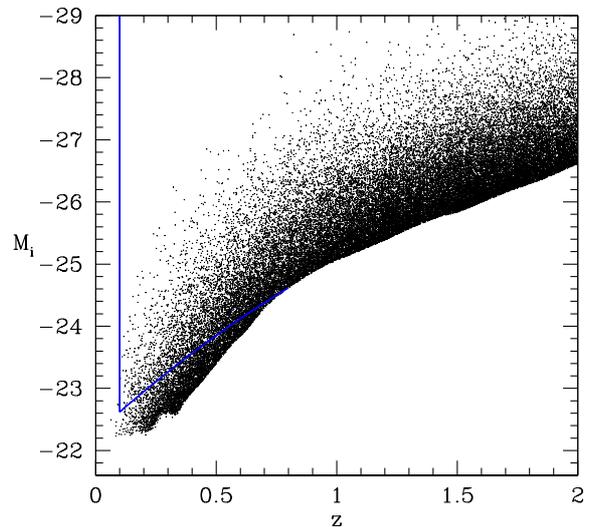}
\caption{Distribution of the uniform-sample quasars in the redshift versus 
$i$-band absolute magnitude plane. The blue line is a low-redshift absolute 
magnitude cut used in our study making the sample nearly constant in 
comoving number density between $z=0.1$ and 1.8.}
\end{figure}

\begin{figure}
\centering
\includegraphics[width=80mm]{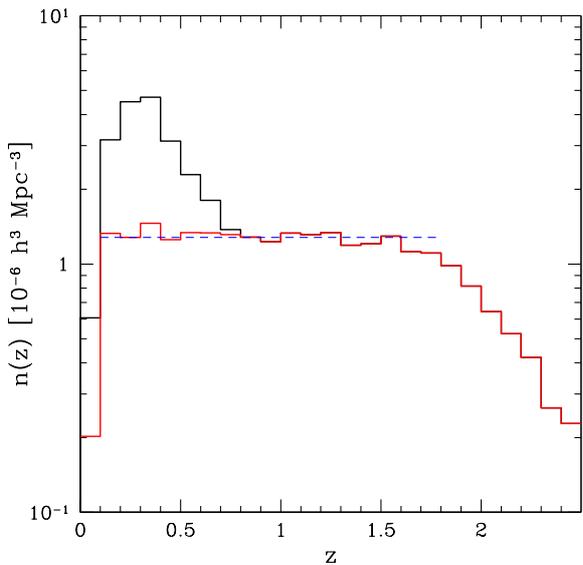}
\caption{Comoving space number density of quasars. The black line is for the whole 
uniform sample, and the red line is when the low-redshift absolute magnitude 
cut (Eq. 1) is applied. The blue dashed line is the mean density between 
$z=0.1$ and 1.8.}
\end{figure}

Our final sample contains $32276$ quasars with $i<19.1$ in the redshift range 
$0.1\le z\le1.8$ covering $5758\textrm{deg}^2$. On the other hand,
\citet{clowes-etal2013} and \citet{nadathur2013}
did not use the sample made to meet the precondition of uniformity
for statistical works. They just applied an $i$-band apparent magnitude cut 
of $i\le19.1$ to \citet{schneider-etal2010}'s sample, where $i$ here is not
corrected for the Galactic extinction. This made their sample affected by
non-uniform quasar target selection which \citet{shen-etal2011} corrected,
and the structures identified are affected by
the Galactic extinction. Our sample is from the homogeneous sample and 
has a true flux limit.
It is also much larger in the sky area (5758 $\textrm{deg}^2$ versus 3725 
$\textrm{deg}^2$) and
in redshift range ($0.1\le z \le1.8$ versus $1.0\le z \le 1.8$).
It makes our sample contain 1.72 times more quasars than that used by
\citet{clowes-etal2013}, \citet{nadathur2013}, and \citet{einasto-etal2014}.

\section{Identification of Quasar Groups}
To identify groups of quasars we apply the Friend-of-Friend (FoF) algorithm
to the `uniform' constant-number density sample described in the previous 
section. We first use LL$=70 h^{-1}$Mpc for a comparison 
with \citet{clowes-etal2013}'s results. 
This corresponds to the threshold overdensity of 
$\delta=(\bar{d}/LL)^3-1=(91.5/70)^3-1=1.23$,
where $\bar{d}$ is the mean quasar separation.
We find 3233 quasar groups having 3 or more members.
Figure 4 shows the positions of top ten richest groups on the sky (upper panel)
and in the $x$-$y$ plane of Equatorial coordinate frame (bottom panel). 
The second richest one labelled with `2' has 70 member quasars and the maximum 
extent of 703 $h^{-1}$Mpc.   
The maximum extent is defined as the comoving distance between the most 
distant two members in a group.
This group corresponds to huge Large Quasar
Group, U1.27, reported by Clowes et al.
The group identified by us has 3 new members and does not have 6 members
compared to Clowes et al.'s list.
We discover a rich quasar group, marked with `1' in Fig. 4, that contains
95 members and the maximum extent of 712 $h^{-1}$Mpc.
This group was not found by previous studies because its center is located 
at $z=0.366$, which is
outside the redshift limit of $1.0<z<1.8$ adopted by Clowes et al.,
\citet{nadathur2013}, and \citet{einasto-etal2014}. 
This group is much richer and has a size larger than U1.27.
A catalog of twenty richest quasar groups is given in Table 1 for future
analyses.

\begin{table}[t!]
\caption{The twenty richest SDSS DR7 quasar groups identified with the linking length of 70 $h^{-1}$Mpc
or overdensity threshold of 1.23. \label{tab:jkastable1}}
\centering
\begin{tabular}{ccccccc}
\toprule
ID     &  ${\bar \alpha}$(deg)$^{\rm a}$   &  ${\bar \delta}$(deg)$^{\rm a}$    &  ${\bar z}$$^{\rm a}$     & ${\bar r}$$^{\rm a}$ &
          N$^{\rm b}$                      &  L$^{\rm c}$ \\
\midrule
     1 & 170.08  & 29.33  & 0.366 & 1012   & 95     & 712 \\
     2 & 163.53  & 14.47  & 1.244 & 2757   & 70     & 703 \\
     3 & 196.45  & 39.95  & 1.112 & 2546   & 64     & 536 \\
     4 & 195.98  & 27.09  & 1.554 & 3198   & 62     & 625 \\
     5 & 166.91  & 33.88  & 1.075 & 2485   & 60     & 570 \\
     6 & 156.69  & 38.63  & 0.580 & 1519   & 59     & 536 \\
     7 & 219.41  & 27.68  & 0.598 & 1560   & 56     & 459 \\
     8 & 204.40  & 13.04  & 1.215 & 2712   & 55     & 582 \\
     9 & 187.60  & 43.55  & 1.404 & 2995   & 53     & 651 \\
    10 & 232.82  & 23.19  & 1.511 & 3141   & 53     & 556 \\
    11 & 219.69  & 19.68  & 0.764 & 1909   & 53     & 640 \\
    12 & 226.57  & 16.83  & 1.062 & 2463   & 51     & 470 \\
    13 & 141.58  & 47.90  & 1.157 & 2620   & 50     & 526 \\
    14 & 208.63  & 25.81  & 1.236 & 2745   & 48     & 446 \\
    15 & 137.82  & 21.70  & 0.957 & 2279   & 47     & 414 \\
    16 & 212.06  & 28.23  & 1.080 & 2493   & 47     & 754 \\
    17 & 127.45  & 20.13  & 1.351 & 2917   & 46     & 545 \\
    18 & 157.40  & 22.01  & 1.493 & 3117   & 45     & 576 \\
    19 & 158.17  & 15.82  & 0.744 & 1869   & 45     & 456 \\
    20 & 231.00  & 47.84  & 1.508 & 3138   & 44     & 459 \\
\bottomrule
\end{tabular}
\tabnote{
The full catalog with information on members of groups can be downloaded at
http://astro.kias.re.kr/Quasar/DR7\_20groups/. \\
$^{\rm a}$ Coordinates of the geometric center of groups. 
The mean comoving distance ${\bar r}$ is in units of $h^{-1}$Mpc. \\
$^{\rm b}$ The number of member quasars. \\
$^{\rm c}$ The maximum extent of groups in $h^{-1}$Mpc.}
\end{table}

To check if our group finding is consistent with previous works we perform the FoF 
group finding to a sample from the original \citet{schneider-etal2010}'s catalog
that is limited only by $i<19.1$ not corrected for Galactic extinction
and a redshift range of $1.0<z<1.8$. In this case 
the richest group is found to have 73 members that are exactly the same as 
the member quasars of U1.27 listed in Table 1 of Clowes et al.
This proves that the difference in our quasar catalog
is not due to the group finding algorithm but to the sample definition.
We also compared our quasar groups with those identified by \citet[see their Table A.3]{einasto-etal2014}
who also used Schneider et al.'s catalog 
constrained by $i<19.1$, and found a very good agreement at $z>1.0$.
We choose our sample instead of that used by previous works
because the observed sample must be uniformly defined 
for statistical hypothesis tests against random catalogs.
The observed quasar distribution has random perturbations along
the line-of-sight due to the random redshift error which is typically about 0.004
in the case of SDSS DR7 catalog \citep{schneider-etal2010}.
Therefore, it is necessary to examine if the quasar group finding is significantly
affected by this error. We will discuss it in the next section.

\begin{figure}
\centering
\includegraphics[width=80mm]{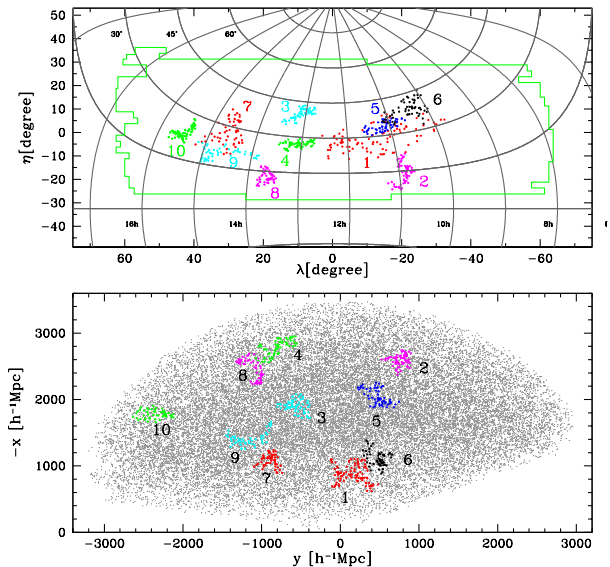}
\caption{
(upper panel) Ten richest quasar groups detected with the linking length 
of $70 h^{-1}$ Mpc. 
The second richest one in magenta on the bottom right labelled with 2
corresponds to the huge Large Quasar Group, U1.27, of \citet{clowes-etal2013}. 
(bottom panel) Same as the upper plot but in the Cartesian coordinates defined 
in the Equatorial coordinate system. Grey points are all the quasars in our sample.}
\end{figure}

\section{Comparison with random distributions}
We generate random point sets which have the same number 
density and the same angular coverage on the sky as our
quasar sample. 
The random catalogs also have the mean point separation of $91.5h^{-1}$Mpc 
throughout the entire redshift range from $z=0.1$ to $1.8$.
We make 10000 random mock samples. For each mock sample we find groups 
of points in the same way as for the quasar sample.
We then measure richness and size of the groups in quasar sample and random 
mock samples.
Richness is the number of member quasars/points belonging to a given group.
Size is the maximum comoving extent of the group.

\begin{figure}
\centering
\includegraphics[width=84mm]{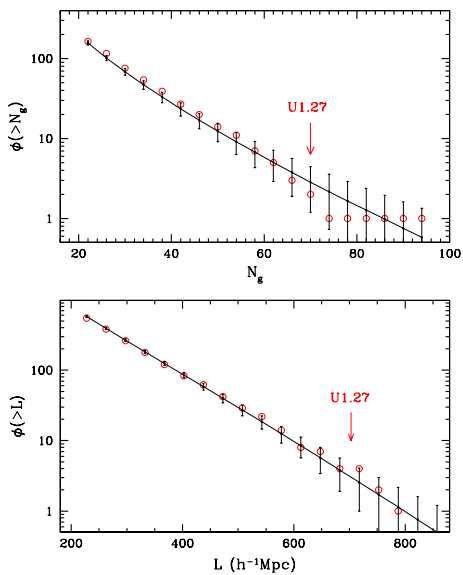}
\caption{Cumulative distribution functions of richness (upper panel) and
maximum extent (bottom panel) of the groups identified in the observed 
SDSS quasar catalog (red circles) and in random point catalogs 
(solid line with error bars).
A set of 10000 random point catalogs is used for the comparison.
The  group corresponding to the huge Large Quasar Group, U1.27, 
reported by \citet{clowes-etal2013} is marked with downward arrows.}
\end{figure}

In the upper panel of Figure 5 we plot the cumulative distribution 
function (CDF) of richness.
It is the number of groups whose richness is greater than or equal to $N_g$.
The circles are from the observed quasar sample. 
Richness of the group corresponding to U1.27 group is 70, and is marked 
with an arrow.  The solid line and error bars 
are the mean and standard deviations obtained from 10000 random catalogs.
Similarly, the bottom panel shows the number of observed quasar groups 
whose size is greater than or equal to L, and the results from 10000 mock catalogs.
We find that U1.27 is 703 $h^{-1}$Mpc long (marked with an arrow) and 
is only the fourth largest one.
It can be seen that the richness and size distributions of the observed quasars 
are statistically indistinguishable with those of random catalogs.

To estimate the probability of finding groups like the observed quasar groups
in random catalogs we perform a $\chi^2$ test.
We define
\begin{equation}
\chi^2 = {1 \over N}\sum_{i=1}^{N} (\phi_i - {\bar \phi}_i)^2/\sigma_i^2
\end{equation}
as a measure of the deviation of a particular CDF $\phi$
from the average CDF ${\bar \phi}$ of random catalogs. 
Here $\sigma_i$ is the standard deviation of the CDF's of random catalogs
in the $i$-th bin,
and the summation is over the range where the average CDF of
random catalogs is between 3 and 100. The choice ensures that
a large part of CDF is compared.
The histograms in Figure 6 are the distributions of
$\chi^2$ of 10000 random catalogs and the arrows correspond to the observed
SDSS quasar catalog generated with LL$=70 h^{-1}$Mpc. 
The probability of having $\chi^2$ greater than $\chi^2_{\rm obs}$ 
in random catalogs is 58.1\% and 92.4\% for richness and size 
distributions, respectively.
Therefore, it can be concluded that the quasar groups identified with 
LL$=70 h^{-1}$Mpc are consistent with those of random point distributions.

On the other hand, we find that
the probability of finding a random group whose richness is equal to or greater than 
$70$ is 92.3\%.  Similarly,
the probability of finding a random group with the maximum extent equal to 
or greater than 703 $h^{-1}$Mpc
is 100\%. 
Therefore, both richness and size of the objects like U1.27
are of no surprise even in random distributions of points with no physical
clustering when LL is as large as $70 h^{-1}$Mpc.

\begin{figure}
\centering
\includegraphics[width=84mm]{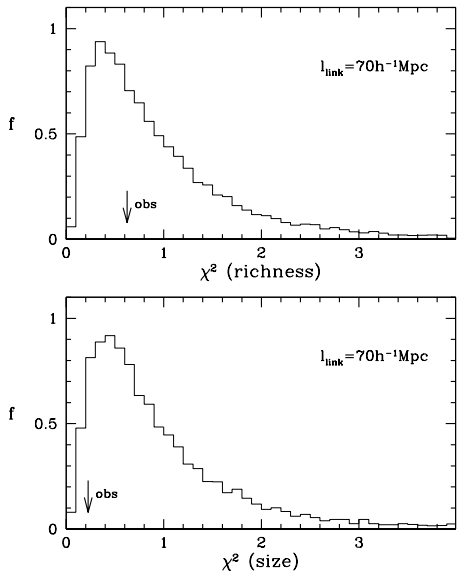}
\caption{Distributions of $\chi^2$ defined as in Eq. (2)  for random point samples. 
The upper panel is for richness of groups identified with the linking length
of 70 $h^{-1}$Mpc, and the bottom panel is for the maximum extent of groups.
The $\chi^2$ for the observed sample are marked with arrows.}
\end{figure}

Since quasars do have intrinsic spatial clustering, it is expected that these 
distribution functions are statistically different from those of random catalogs
for some smaller LL or for higher density thresholds.
Figure 7 shows the CDF's of richness and size of quasar
groups when LL is taken to be 60, 50, and 40 $h^{-1}$Mpc.
Also shown are the average CDF's (solid lines with error
bars) of random point groups identified with the corresponding LL.
It is clear that the CDF's of observed group richness and size 
become more different from those of random catalogs as LL
decreases. 

It will be then interesting to search for the critical LL for which 
the observed CDF becomes significantly different from those of random points.
We calculate the probability that the CDF of random point groups
happens to show a deviation from the mean CDF as large as or greater than 
the difference of the observed CDF from the mean.
This is estimated by counting the cases with $\chi^2_{\rm ran} \ge \chi^2_{\rm obs}$.
A high probability means that the observed groups are statistically
consistent with random ones. Figure 8 shows the probability as a function 
of the LL used to find groups. For both richness and size 
it can be seen that the observed groups are consistent with those of 
random catalogs when LL is chosen to be larger than 
about $50 h^{-1}$Mpc.
Clowes et al. chose LL$=70 h^{-1}$Mpc and it is now clear that
the choice makes the group catalog dominated by randomly connected groups.
Figures 7 and 8 indicate that the groups identified with LL$< 50 h^{-1}$Mpc 
are likely to be physically-clustered genuine groups of quasars. 

As mentioned in the previous section the random redshift error of order of 0.004 seems problematic
in finding physical structures as the error corresponds to $\sim12h^{-1}\textrm{Mpc}$.
We use the redshifts calculated by \citet[hereafter HW10]{hewett_wild2010} whose redshift errors
are much smaller and redo the group findings. According to the error estimates for the redshifts of HW10,
the RMS redshift error for quasars at $0.1<z<1.8$ is found to be only 0.00024, more than 10 times
smaller than the nominal value for the SDSS DR7.

It will be a good consistency check to compare two group catalogs derived from \citet{schneider-etal2010} and
HW10's redshifts. We find that the new quasar group catalog is very close to the previous one obtained
using \citet{schneider-etal2010}'s redshifts. All major groups are detected again and only a few members are affected.
As a result, the richness and size distributions change only slightly. It is reassuring that
the quasar groups and the statistics are robust against the redshift error of \citet{schneider-etal2010}.
When $LL=50h^{-1}\textrm{Mpc}$, we still find a very good agreement of both distribution functions.
Therefore, all of our results are essentially unchanged by using more accurate quasar redshifts.
We stick to \citet{schneider-etal2010}'s redshifts in our paper since there is practically no change in the conclusions
and it is easier to compare our groups directly to those of previous works.
The groups identified in the redshift space are also affected by the peculiar velocities of quasars.
Influence of peculiar velocity on group identification is to some extent unavoidable.
But this is relatively small because the RMS pairwise peculiar velocity of galaxies is only about $300km/s$.

\begin{figure}
\centering
\includegraphics[width=84mm]{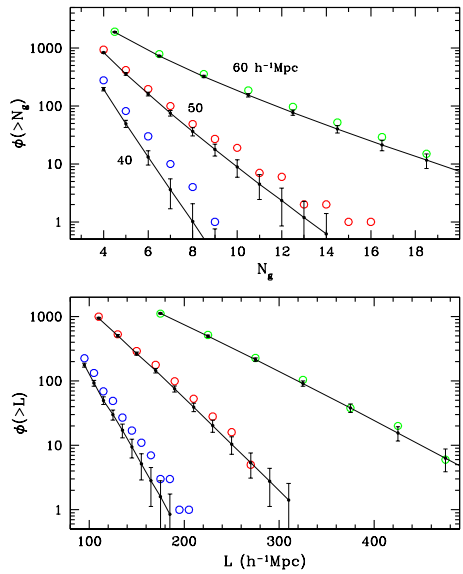}
\caption{Cumulative distribution functions of richness (upper panel) and size (bottom panel)
when the linking length for group identification is taken to be
60, 50, and 40 $h^{-1}$Mpcs. Circles are from observation, and solid lines
with error bars are from 1000 random catalogs.}
\end{figure}

\begin{figure}
\centering
\includegraphics[width=80mm]{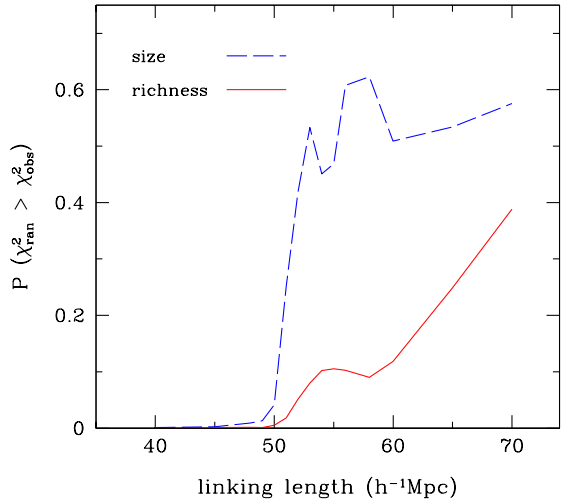}
\caption{Probability of finding random catalogs whose distribution function
deviates from their mean more than or equal to that of the observed sample.
The deviation is measured by $\chi^2$ defined in Eq. (2).
Probability is measured for richness (solid line) and size (dashed line)
distribution functions as a function the linking length.}
\end{figure}

\section{Summary and Conclusions}
We make a wide contiguous volume-limited sample of quasars 
statistically uniformly selected from the SDSS DR7
quasar catalog and identify quasar groups using the FoF algorithm.
The statistical properties of the observed quasar groups are compared with 
those of random point sets. We make the following findings.

1. The observed SDSS DR7 quasar groups identified with 
LL$=70 h^{-1}$Mpc are statistically consistent, in terms of size and richness,
with those of random points clustered by chance.
The size or richness distribution functions of those quasar groups 
can be found in random catalogs with 
92.4\% and 58.1\% probabilities, respectively. 

2. The probabilities of finding groups larger or richer than
the large quasar group U1.27, reported by \citet{clowes-etal2013}, 
in a random point catalog are 100\% and 92\%, respectively.

3. The quasar groups are statistically dominated by physically clustered
quasars when LL for group identification is less than 50 $h^{-1}$Mpc.

We emphasize that the observed LSS should not be used to draw cosmological 
conclusions without making elaborate statistical tests.
Size and richness of cosmic structures depend sensitively on how 
those structures are identified.
For example, one can always find structures as large as the survey size 
if the threshold density level used to identify them is lowered close to
the percolation density. Even in a random point distribution without any
physical clustering an infinitely long structure can be identified
when LL is increased to 0.864 times the mean particle
separation, which is the critical value for percolation.
It is conceivable that one would claim discovery of groups of galaxies
or quasars even larger than the Sloan Great Wall in existing or
future redshift surveys. 
The lesson of the past is that one will again need to make
statistical tests before drawing conclusions on cosmological implication
of those objects.

Our results suggest that the large quasar groups identified with 
LL $> 50 h^{-1}$Mpc can be considered as associations of smaller 
physically-clustered quasar systems that are connected by a very large linking length.
The smaller quasar systems are similar to the supercluster complexes 
in the local universe, as was suggested in \citet{einasto-etal2014}. 
\citet{hutsemekers-etal2014} showed that quasar spin axes in U1.27 seem to be aligned parallel to this structure.
However, from their Figure 6 one can see that alignment is better if this quasar group is divided into smaller systems,
in agreement with our interpretation of quasar systems.

There have also been several studies trying to define and measure the
homogeneity scale of the universe \citep{yadav-etal2010, 
marinoni-etal2012,scrimgeour-etal2012}. However, 
definition of the homogeneity scale has been either arbitary or
dependent on observational samples.
We point out that, in the current popular cosmological models that 
adopt the Cosmological Principle, the notion of the critical scale of 
homogeneity does not make sense.
Homogeneity is only achieved asymptotically in these models as the scale
of observation increases, and there is no characteristic scale
intrinsic to the universe above which the universe can be suddenly regarded 
homogeneous \citep[see also][]{nadathur2013}.
Instead of looking for arbitrarily-defined homogeneity scales
or naively examining the size of the largest structures
the observed LSS should be statistically compared with the structures 
identified in the same way in the mock survey samples obtained from simulated
initially-homogeneous universes.
In our forthcoming paper we will use mock quasars formed in simulated universes
to identify large groups of quasars and statistically compared their properties
with those of observed quasars.
We will examine if the large structures can be explained in the framework
of the standard hierarchical structure formation model
in the given Hubble time.


\acknowledgments
The authors thank the anonymous referee for helpful comments and reviewing our paper quickly, 
and also thank Korea Institute for Advanced Study for providing computing resources
(KIAS Center for Advanced Computation Linux Cluster System).
M. Einasto was supported by the ETAG project IUT26-2, and by the European 
Structural Funds grant for the Centre of Excellence ``Dark Matter 
in (Astro)particle Physics and Cosmology" TK120.
H. Lietzen acknowledges financial support from the Spanish Ministry 
of Economy and Competitiveness (MINECO) under the 2011 Severo Ochoa Program 
MINECO SEV-2011-0187.

Funding for the SDSS and SDSS-II has been provided by the Alfred P. Sloan
Foundation, the Participating Institutions, the National Science
Foundation, the U.S. Department of Energy, the National Aeronautics and
Space Administration, the Japanese Monbukagakusho, the Max Planck
Society, and the Higher Education Funding Council for England.
The SDSS Web Site is http://www.sdss.org/.


\end{document}